\documentclass{ws-procs9x6}

\def\eqnstart{\begin{eqnarray}}
\def\eqnend{\end{eqnarray}}

\begin{document}

\title{Coulomb screening effect on the nuclear-pasta structure}
\medskip

\author{Toshiki Maruyama$^1$, 
Toshitaka Tatsumi$^2$,}
\author{ 
Dmitri N. Voskresensky$^3$,
Tomonori Tanigawa$^{4,1}$,}
\author{Satoshi Chiba$^1$,
Tomoyuki Maruyama$^5$}

\address{}

\address{
1 Advanced Science Research Center, Japan Atomic Energy Research Institute,
Tokai, Ibaraki 319-1195, Japan}
\address{
2 Department of Physics, Kyoto University, Kyoto 606-8502, Japan}
\address{
3 Moscow Institute for Physics and Engineering,
Kashirskoe sh.~31, Moscow 115409, Russia}
\address{
4 Japan Society for the Promotion of Science, Tokyo 102-8471, Japan}
\address{
5 BRS, Nihon University, Fujisawa, Kanagawa 252-8510, Japan}

\maketitle

\abstracts{
Using the density functional theory (DFT) with the relativistic mean 
field (RMF) model,
we study the non-uniform state of nuclear matter, ``nuclear pasta''.
We self-consistently
include the Coulomb interaction together with other interactions.
It is found that the Coulomb screening effect is significant for each 
pasta structure 
but not for the bulk equation of state (EOS) of the nuclear pasta phase.
}

\section{Introduction}

One of the most interesting features of low-density nuclear matter
is the possibility of the
existence of non-uniform structures, called ``nuclear pastas''.\cite{Rav83}
At low densities, nuclei in matter 
are expected to form 
the Coulomb lattice embedded in the neutron-electron seas, so as 
to minimize the Coulomb interaction energy. On the other hand,
another possibility has been discussed:  
the stable nuclear shape may change from sphere to
cylinder, slab, cylindrical hole,
and to spherical hole with increase of the matter density, and 
``pastas'' are eventually dissolved into uniform matter at 
a certain nucleon density close
to the saturation density, $\rho_s\simeq 0.16~$fm$^{-3}$.
The existence of such ``pasta'' phases, instead of the ordinary
crystalline lattice
of nuclei, would 
modify several important processes
in supernova explosions by changing the hydrodynamic properties
and the neutrino opacity in the supernova matter.
Also expected is the influence of the ``pasta'' phases
on star quakes of neutron stars
and pulsar glitches 
via the change of mechanical properties of the crust matter.

Several authors have investigated the low-density 
nuclear matter using various models.
$^{1-8}$
Roughly speaking, the favorable nuclear shape is determined by 
a balance between the
surface and the Coulomb energies, as has been shown by 
previous studies, 
where the rearrangement effect on the density profile of the
charged particles, especially electrons, by the Coulomb interaction is
discarded. However, 
the proper treatment of the Coulomb interaction should be very important, as
it is demonstrated in 
Ref.~9;
the screening of the Coulomb interaction by the charged particles may give a
large effect on the stability of the geometrical structures.

We have been recently exploring the effect of the Coulomb screening in the 
context of the structured mixed phases in various first order phase
transitions such as hadron-quark
deconfinement transition, kaon condensation and liquid-gas transition
in nuclear matter. We treat the nuclear ``pasta'' phases as a part of
our project, since they can be considered as 
structured mixed phases during the liquid-gas transition in nuclear matter.

Our aim here is to study the nuclear ``pasta'' structures by means of
a mean field model, which includes the Coulomb interaction in a proper  
way, and we figure out how the Coulomb screening effect modifies
the previous results without it.

\section{Density Functional Theory with the Relativistic Mean-field Model}

To study the non-uniform nuclear matter,
we follow the  density functional theory (DFT) with
the relativistic mean field (RMF) model.\cite{refDFT}
The RMF model with fields of mesons and baryons 
is rather simple 
for numerical 
calculations, but realistic enough to reproduce 
main nuclear matter properties.
In our framework, the Coulomb interaction is properly included in the 
equations of motion for nucleons, electrons and the meson mean fields,
and we solve the Poisson equation for the Coulomb potential $V_{\rm
Coul}$ 
self-consistently with them.
Thus the baryon and electron density profiles, as well as the meson
mean fields, are determined in a way fully
consistent with the Coulomb potential.
Note that our framework can be easily extended to other situations. For
example, 
if we 
take into account meson condensations, which are likely realized in a high-density
region, we should
only add the relevant  meson field terms. 
In Ref.~11
we have included 
the kaon degree of freedom in such a treatment to discuss kaon condensation 
in high density regime.

To begin with, we present the thermodynamic potential of the form,
\begin{eqnarray}
\Omega&\!=\!&\Omega_B+\Omega_M
    +\Omega_e,\\
\Omega_B&\!=\!&\int\!\!\! d^3r\!\left[
  \sum_{i=p,n}\left({2\over(2\pi)^3}
  \int_0^{k_{Fi}}d^3k \sqrt{{m_B^*}^2+k^2}-\rho_i\nu_i\right)
\right],\\
\Omega_M&\!=\!&\int\!\!\! d^3r\!\left[\!
  {(\nabla\sigma)^2\!+\!m_\sigma^2\sigma^2 \over2}\!+\!U(\sigma)
  \!-\!{(\nabla\omega_0)^2\!+\!m_\omega^2\omega_0^2 \over2}
  \!-\!{(\nabla\rho_0)^2\!+\!m_\rho^2\rho_0^2\over2} \!\right]\!\!,\\
\Omega_e&\!=\!&\int\!\!\! d^3r\!\left[
-{1\over8\pi e^2}(\nabla {V_{\rm Coul}})^2-{({V_{\rm Coul}}-\mu_e)^4\over12\pi^2}
\right],
\end{eqnarray}
where
$\nu_p=\mu_B-\mu_e+{V_{\rm Coul}}-g_{\omega N}\omega_0-g_{\rho N}\rho_0,\ \ 
\nu_n=\mu_B-g_{\omega N}\omega_0+g_{\rho N}\rho_0,\ \ 
m_B^*=m_B-g_{\sigma N}\sigma,\ \ 
U(\sigma)={1\over3}bm_B(g_{\sigma N}\sigma)^3+{1\over4}c(g_{\sigma N}\sigma)^4.
$
Here we used the local-density approximation for nucleons and electrons,
while one still should carefully check  its validity. 
The introduction of the
density variable 
is meaningful, if the typical length 
of the nucleon density variation  inside the structure
is larger than the
inter-nucleon distance, which we assume to be fulfilled. 
We must also keep in mind that the approximation
is broken down for small
structure sizes, since quantum effects become prominent there.
For the sake of simplicity
we also  omitted nucleon and electron density derivative terms.
If the nucleon length scale were shorter than 
lengths of changes of the meson mean fields, one could not 
introduce
the derivatives of the nucleon density but could simplify the problem 
introducing the corresponding contribution to the 
surface tension.
In the given case (when we suppressed the derivative terms mentioned above) 
the resulting nucleon density
follows the changes of the meson mean fields. 
However, even in this case the presence
of the derivative terms (of the same order as for other fields)
could affect the numerical
results. 
Here we consider large-size pasta structures and 
simply discard the density variation effect, as a first-step calculation,
while it can be incorporated in the quasi classical
way by  the derivative expansion within 
the density functional theory.\cite{refDFT}  
The parameters are set to reproduce the nuclear-matter saturation properties.
 From the variational principle, 
 ${\delta\Omega\over\delta\phi_i({\bf r})}=0$ 
($\phi_i=\sigma,\rho_0,\omega_0,V_{\rm Coul}$) or 
${\delta\Omega\over\delta\rho_i({\bf r})}=0 (i=n,p,e)$,
we get the
coupled equations of motion as
\begin{eqnarray}
-\nabla^2\sigma+m_\sigma^2\sigma &=& -{dU\over d\sigma}+
g_{\sigma N}(\rho_n^{(s)}+\rho_p^{(s)})
    \\
-\nabla^2\omega_0+m_\omega^2\omega_0 &=& g_{\omega N}(\rho_p+\rho_n)
    \\
-\nabla^2\rho_0+m_\rho^2\rho_0 &=& g_{\rho N}(\rho_p-\rho_n)
    \\
\nabla^2{V_{\rm Coul}} &=& 4\pi e^2{\rho_{\rm ch}} \ \ \ \ 
(\hbox{charge density}\ \ {\rho_{\rm ch}}\;=\;{\rho_p}+{\rho_e}
     )\label{poisson}\\
\mu_n=\mu_B &=& \sqrt{k_{Fn}^2+{m_B^*}^2}+g_{\omega N}\omega_0-g_{\rho N}
\rho_0\\
\mu_p=\mu_B-\mu_e &=& \sqrt{k_{Fp}^2+{m_B^*}^2}+g_{\omega N}\omega_0+
g_{\rho N}\rho_0-{V_{\rm Coul}}\\
{\rho_e}&=&-(\mu_e-{V_{\rm Coul}})^3/3\pi^2.
\end{eqnarray}
Note that first, the Poisson equation (\ref{poisson}) is a highly nonlinear 
equation for
$V_{\rm Coul}$, since  $\rho_{\rm ch}$ in r.h.s. includes it in a
complicated way, and secondly, the Coulomb potential always enters the
equation through the gauge invariant combination, $\mu_e-V_{\rm Coul}$.

To solve the above coupled equations numerically,
we use the Wigner-Seitz cell approximation:
the space is divided into equivalent cells 
with some geometrical symmetry.
The shape of the cell changes: 
sphere in three dimensional (3D) calculation, cylinder in 2D and slab in 1D, 
respectively. 
%
Each cell is charge-neutral and
all the  physical quantities in a cell are smoothly connected to those  
of the neighbor cell with zero gradients at the boundary.
The cell is divided into grid points ($N_{\rm grid}\approx 100$) and
the differential equations for fields are solved by the relaxation method
with constraints of given baryon number and charge neutrality.

\section{Bulk Property of Finite Nuclei}

Before applying our framework to the problem of the pasta phases in nuclear matter, we check how it 
can describe finite nuclei.
In this calculation, the electron density is set to be zero and 
the boundary condition or the
charge-neutrality condition is not imposed.
However, we assumed  the spherical shape of nuclei.
In Fig.~1 (left panel) we show the density profiles of some typical nuclei.
We can see how well our framework may describe the density profiles
 of these nuclei. 
To get a better fit, especially around the surface region, 
we might need to include the derivative terms, as we have mentioned.
Fine structures seen in the empirical 
density profiles, which come from the shell
effects 
(see, e.g., a proton density  dip at the center of a 
light $^{16}$O nucleus),  
cannot be
reproduced by the mean field approach.
By imposing the beta equilibrium, 
the most stable proton ratio
can be obtained for a given mass number.
Figure 1 (right panel) shows the mass-number dependence of the
binding energy per nucleon and the proton ratio.
We can see that the bulk properties of finite nuclei (density, binding 
energy and
proton ratio) are satisfactorily reproduced for our present purpose.

Note that we should adjust a slightly smaller 
value of the  sigma mass 
than that one usually  uses, i.e.~400 MeV, to get such a good fit.
If we used the popular value of $m_\sigma\approx 500$ MeV, 
finite nuclei would be overbound by about 3 MeV per nucleon.
Although  the actual
value of the  sigma mass (or the omega mass)
has little relevance for infinite nuclear matter, 
it is important for finite 
nuclei and other non-uniform nucleon systems,
since the meson mass characterizes the interaction range 
and consequently affects, e.g., the nuclear surface property. 
\begin{figure}
  \includegraphics[height=.29\textheight]{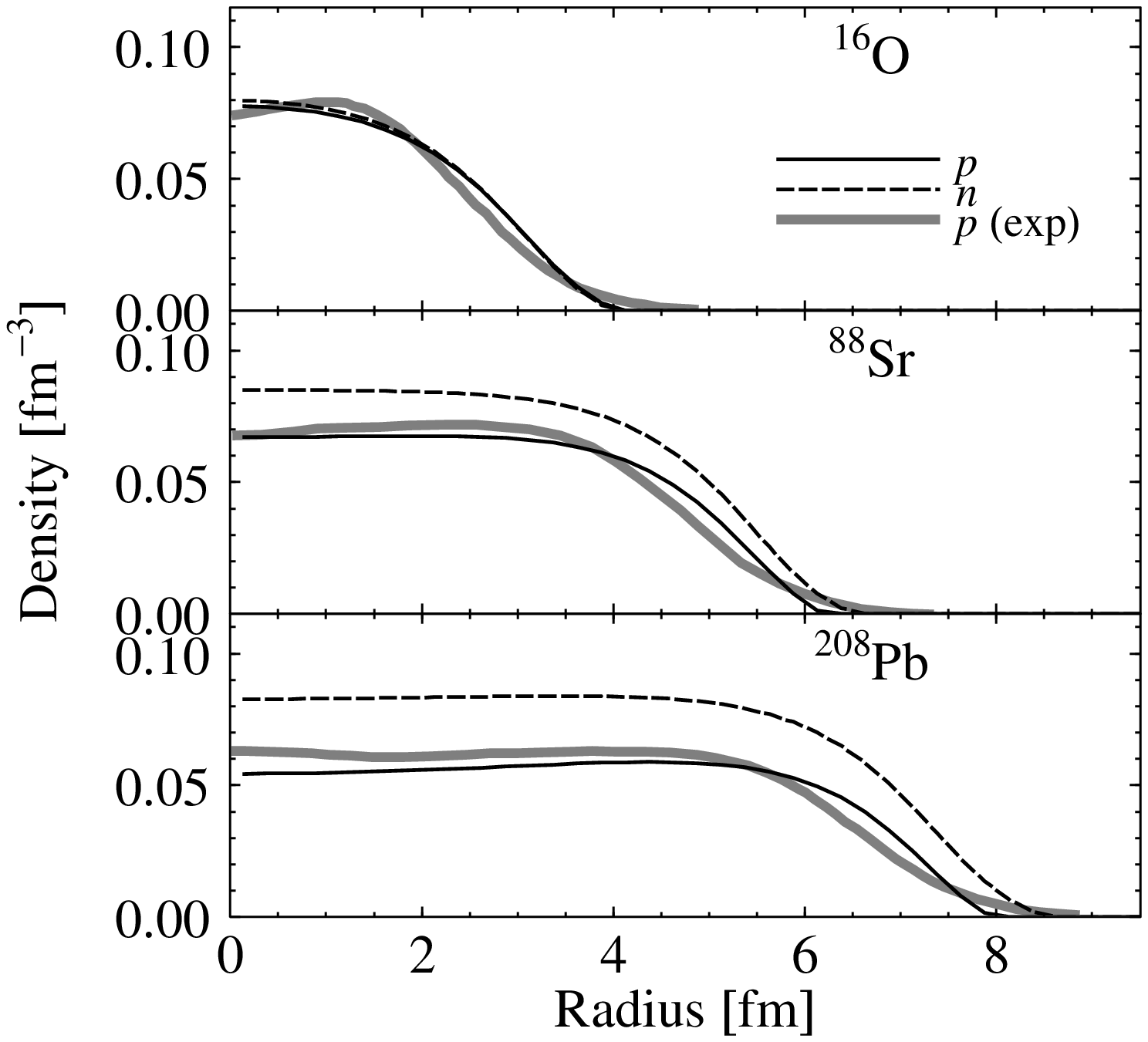}
  \includegraphics[height=.29\textheight]{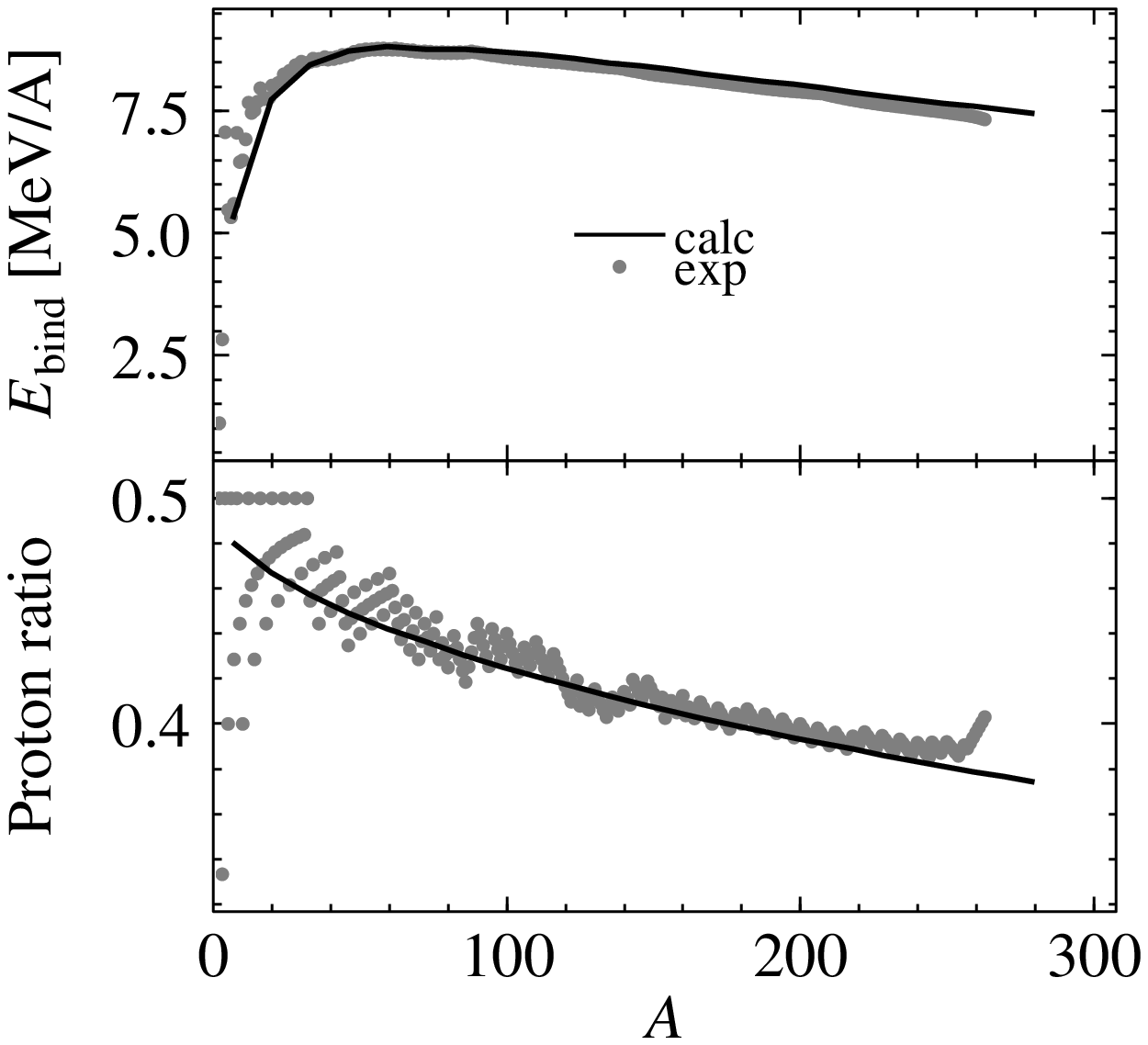}
  \caption{Left: the density profiles of typical nuclei.
The proton densities (solid curves) are compared with the experiment.
Right: the binding energy per nucleon and the proton ratio 
of finite nuclei.
}
\end{figure}

\section{Nuclear ``Pasta'' at Sub-nuclear Densities}

In the density region, where nuclei are 
about to dissolve 
into uniform nuclear matter,
it is expected that the energetically favorable mixed phase,
which consists of nuclei and free nucleons,
possesses interesting geometrical structures,
such as  rod-like and slab-like nuclei and rod-like and spherical bubbles, etc.
These nuclei with exotic geometrical shapes are referred 
to as nuclear ``pastas''.
Phenomenologically the existence of the ``pasta'' phases instead 
of the crystalline lattice
of nuclei would affect the mechanism of 
supernova explosions and the glitch phenomena in pulsars.
Due to these important consequences 
the ``pasta'' structure has been
studied by number of authors.$^{1-8}$
It is widely accepted that the ``pasta'' structure is realized 
basically due to
the balance of the Coulomb energy and the surface tension.
However, the electron density has been always treated as an uniform
background in the standard treatments.
Here we study the nuclear ``pasta'' structure
within our framework, which  consistently treats the Coulomb potential and the 
electron distribution.

\subsection{Symmetric Nuclear Matter and the Coulomb Screening Effect}

First, we focus on symmetric nuclear matter (relevant to 
supernova matter at the initial stage of collapse) where 
the Coulomb screening effect by electrons is expected to be large.
Figure 2 shows some typical density profiles in the Wigner-Seitz cells.
The geometrical dimension of the cell  is denoted as ``3D'' 
(three dimensional), etc.
The horizontal axis in each panel denotes the radial distance 
from the center of the cell,
and the boundary is indicated by the hatch. 
The nuclear ``pasta'' structures are clearly seen. 
Note that the electron density profile becomes no more uniform due to the
Coulomb screening. 

The phase diagram of the matter structure is shown in Fig.~3 (left).
The size of the cell $R_{\rm cell}$ is optimized with precision of 1 fm,
and  the lowest energy configurations are chosen among various
geometrical structures.
In the figure, there never appears the spherical-hole configuration.
This is one of the consequences of the Coulomb screening effect.
It should be noted that appearance of the pasta structures 
is also sensitive to the choice of
the effective interaction, 
as discussed by Oyamatsu et al.\cite{Oyamatsu04}
%
\begin{figure}
  \includegraphics[height=.39\textheight]{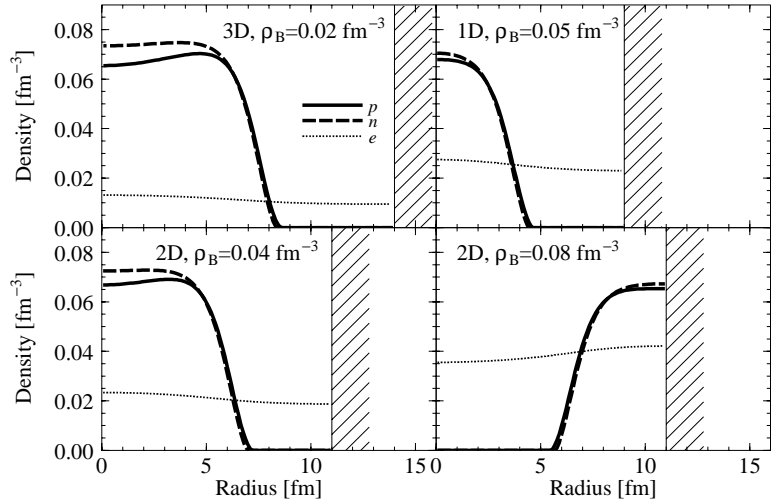}
  \caption{
Examples of the density profiles in the cell for symmetric nuclear matter (droplet, rod, slab, and tube).
}
\end{figure}
\begin{figure}
  \includegraphics[height=.39\textheight]{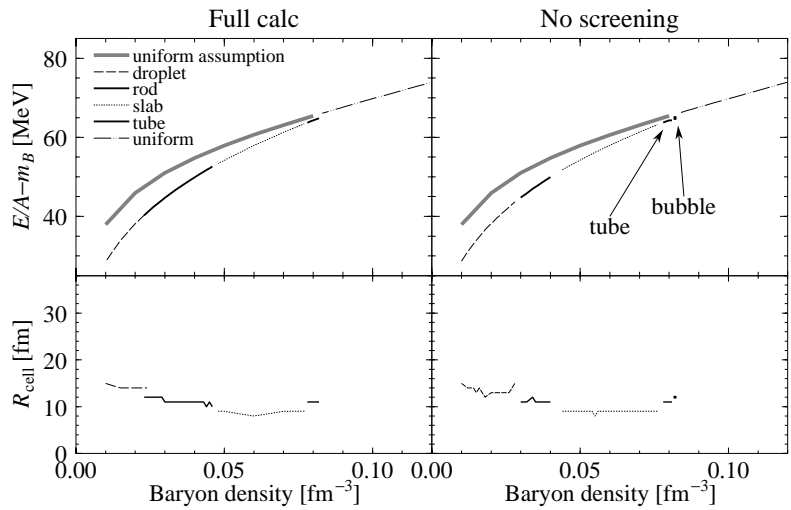}
  \caption{
Left: the binding energy per nucleon and the cell size 
of symmetric nuclear matter.
Right: the same as the left panel with the uniform electron distribution.}
\end{figure}

To elucidate the Coulomb screening effect, there are two possible ways:
one is to compare our results with those given by solving the equations
of motion 
for fields and the density profiles neglecting 
the Coulomb potential $V_{\rm Coul}$. 
Then the Coulomb energy is 
calculated by the use of the density profiles thus determined and
finally added to the total energy, as in the simple bulk calculations.
The optimum cell size is determined by this total energy including
the Coulomb energy.
The other way is to compare our results with those given by only 
discarding $V_{\rm Coul}$ in r.h.s. of the Poisson
equation, while keeping it in other equations of motion. It  
is equivalent to the assumption of the {\it uniform} electron density distribution,
which has been used in the previous studies;$^{3-8}$ 
%
there protons interact with 
each other and may form a non-uniform structure through the balance 
between the nuclear surface tension and the Coulomb interaction 
in a {\it uniform} electron background. 
Thus the density rearrangement effect is
partially taken into account for protons, while it is completely neglected for electrons. 
The first way may be standard to extract the Coulomb effect and to
compare our full calculation with the bulk one; 
actually this way has been taken in the context of kaon condensation\cite{maru}
or hadron-quark deconfinement transition.\cite{voskre}
However, we don't take the first way and dare to take
the second way here  
to compare our results with the previous ones given by the uniform-electron calculation. 

We show in the right panel of Fig.~3 the results without the Coulomb screening 
(uniform-electron calculation). 
The region of each structure (droplet, rod, etc.) is different from that
given by the full calculation. 
Especially, the ``bubble'' (spherical hole) appears in this case.
Since the appearance of various geometrical structures and their 
region depend on
the very subtle energy difference, 
the Coulomb screening has a significant influence on the sequence of the
different pasta phases. 

Comparing the case of uniform matter with the case of the pasta phases, 
one can see that the non-uniform structures reduce the energy. 
However, the Coulomb screening effect on the bulk EOS 
(difference between left and right panels of Fig.~3) 
is rather small.

\subsection{Nuclear Matter in Beta Equilibrium}

\begin{figure}
  \includegraphics[height=.39\textheight]{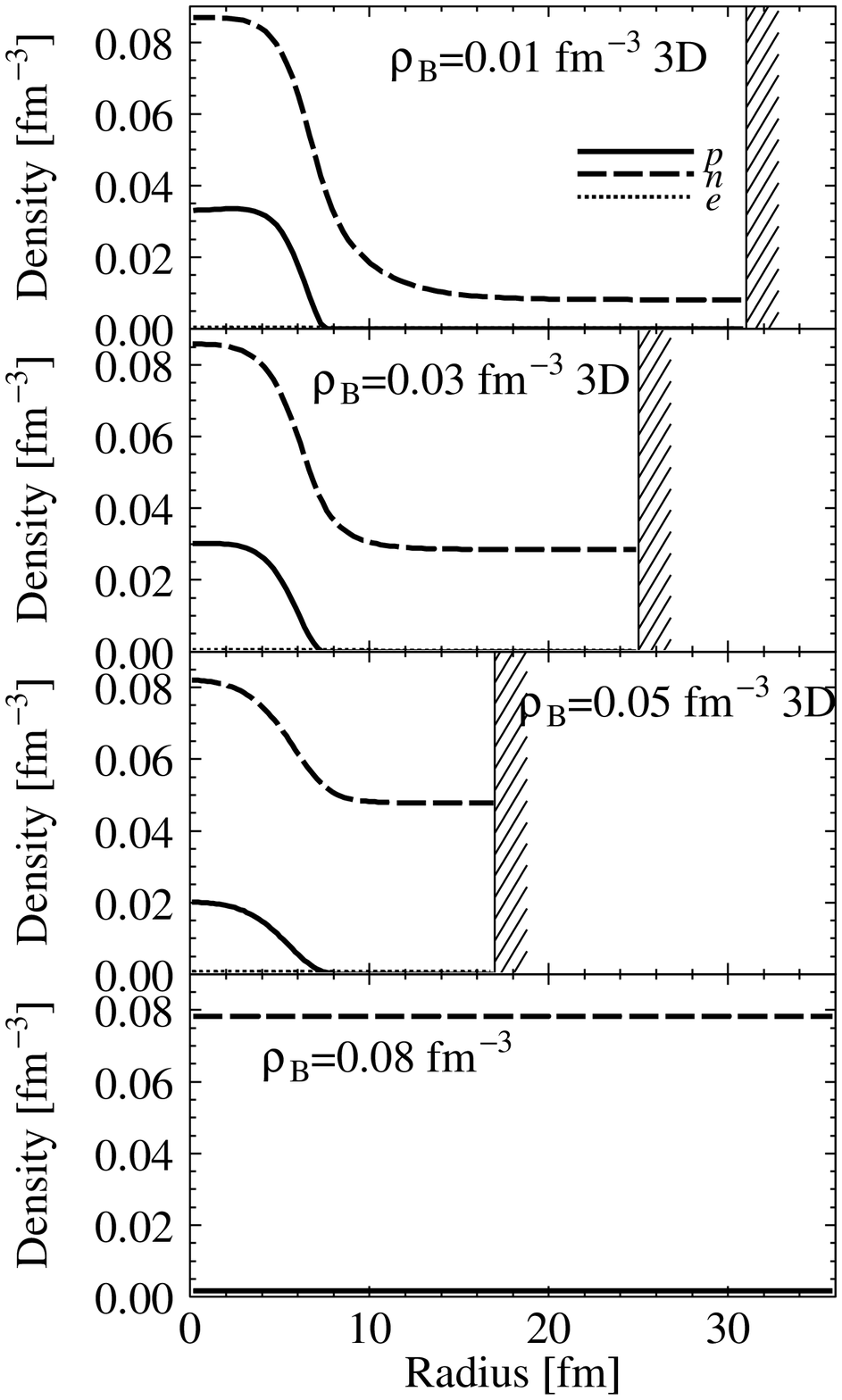}
  \includegraphics[height=.39\textheight]{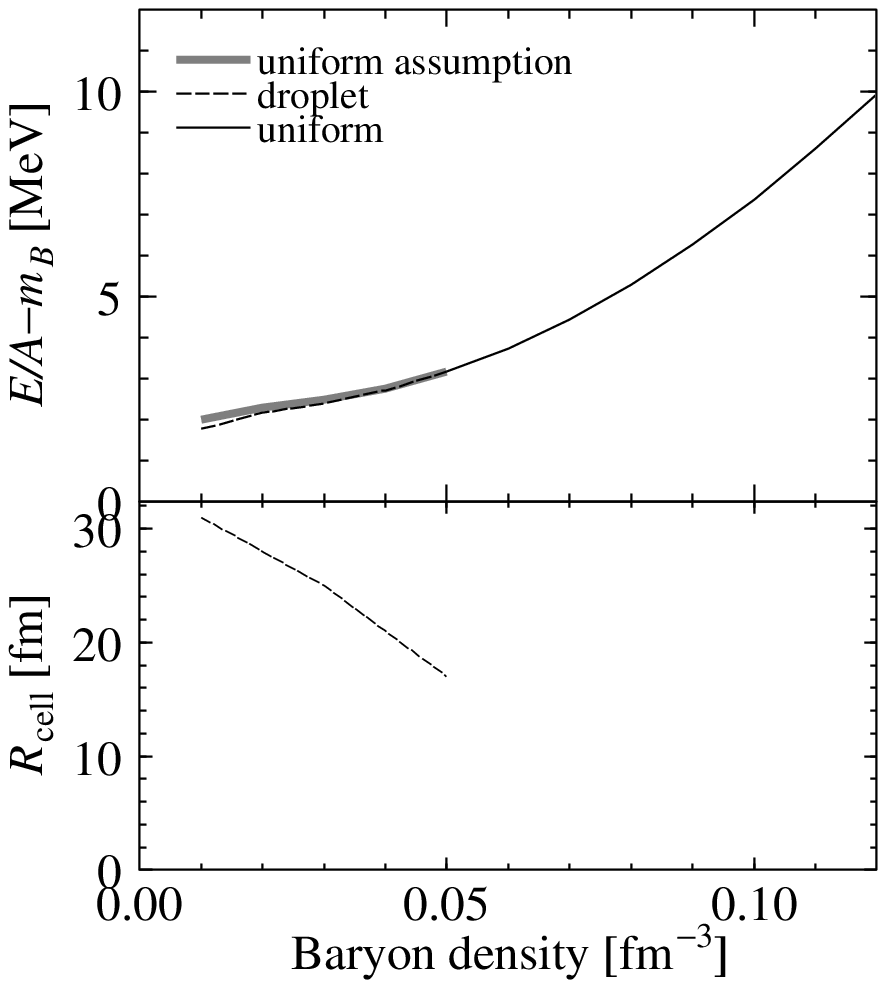}
  \caption{
Left: examples of the density profiles of nuclear matter in beta equilibrium.
The electron density is too small to see clearly.
Right: the binding energy per nucleon and the cell size. 
}
\end{figure}

Next, we discuss the nuclear matter in beta equilibrium, which is
relevant to stable neutron stars.
Figure 4 (left) shows the density profiles at several densities. 
Only the case for three dimensional (3D) geometrical structure is shown, 
since 2D and 1D cases are energetically unfavored in our calculation. 
One can see 
the proton-enriched droplets embedded into the neutron sea 
at low densities.
The EOS with the 3D  phases is shown in the right panel 
of Fig.~4.
In the beta equilibrium nuclear matter the effect of the non-uniform structure 
becomes much less compared to that for symmetric nuclear matter. 
Since the electron fraction is small, the Coulomb screening effect
should not be remarkable.

\section{Summary and Concluding Remarks}

We have discussed the low-density nuclear matter structures 
``nuclear pastas'' and elucidated the Coulomb screening effect.
Using a self-consistent framework based on DFT and RMF, 
we took into account the Coulomb interaction in a proper way and 
numerically solved the coupled equations of motion to extract the
density profiles of nucleons. 

First we have checked how realistic our framework is by calculating the bulk
properties of finite nuclei as well as 
the saturation properties of nuclear matter, and found it can describe
both features satisfactorily. 

In symmetric nuclear matter, we have observed the ``nuclear pasta''
structures with various geometries at sub-nuclear densities. 
The appearance of the pasta structures significantly lowers the energy, 
i.e. softens the EOS, while 
the energy differences between various geometrical structures are
rather small. So the Coulomb screening effect and the rearrangement of the
charge density can affect such changes of geometrical structures in
spite of that 
its absolute value is rather small in comparison with nuclear
interaction energy.

By comparing the results with and without the Coulomb screening,
we have seen that the self-consistent inclusion of the Coulomb
interaction changes the phase diagram. 
In particular one of the pasta configurations appears only 
when the Coulomb screening is switched off in our calculation.
The effects of the Coulomb screening on the EOS, on the other hand,
was found to be small.

We have also studied the structure of nuclear matter in 
beta equilibrium.
There we have observed only proton-enriched droplet in the neutron sea.
No other geometrical structures like rod, slab, etc.~appeared. 

Detailed discussions about the Coulomb screening effect on the nuclear
pasta phases will be reported elsewhere.


\begin{thebibliography}{99}

\bibitem{Rav83} 
D.~G.~Ravenhall, C.~J.~Pethick and J.~R.~Wilson,
{\it Phys. Rev. Lett.} {\bf 27}, 2066 (1983).



\bibitem{Has84} 
M.~Hashimoto, H.~Seki and M.~Yamada,
{\it Prog. Theor. Phys.} {\bf 71}, 320 (1984).

\bibitem{Wil85} 
R.~D.~Williams and S.~E.~Koonin,
{\it Nucl. Phys.} {\bf A435}, 844 (1985).

\bibitem{Oya93} 
K.~Oyamatsu,
{\it Nucl. Phys.} {\bf A561}, 431 (1993).

\bibitem{Lor93} 
C.~P.~Lorentz, D.~G.~Ravenhall and C.~J.~Pethick,
{\it Phys. Rev. Lett.} {\bf 25}, 379 (1993).

\bibitem{Cheng97}
K.~S.~Cheng, C.~C.~Yao and Z.~G.~Dai,
{\it Phys. Rev.} {\bf C55}, 2092 (1997).

\bibitem{Mar98}
T.~Maruyama, K.~Niita, K.~Oyamatsu, T.~Maruyama, S.~Chiba and A.~Iwamoto, 
{\it Phys. Rev.}  {\bf C57}, 655 (1998).
T.~Kido, T.~Maruyama, K.~Niita and S.~Chiba,
{\it Nucl. Phys.} {\bf A663}-{\bf 664}, 877 (2000).

\bibitem{Gen00}
G.~Watanabe, K.~Iida and K.~Sato,
{\it Nucl. Phys.} {\bf A676}, 445 (2000);
G.~Watanabe, K.~Sato, K.~Yasuoka and T.~Ebisuzaki,
{\it Phys. Rev.}  {\bf C66}, 012801 (2002).

\bibitem{voskre} 
D.N.~Voskresensky, M.~Yasuhira and T.~Tatsumi,
{\it Phys. Lett.} {\bf B541}, 93 (2002); {\it Nucl. Phys.}
 {\bf A723}, 291 (2003); T. Tatsumi and D.~N. Voskresensky, this
	proceedings (nucl-th/0312114). 

\bibitem{refDFT} 
{\it Density Functional Theory}, ed. E.~K.~U.~Gross and R.~M.~Dreizler, 
Plenum Press (1995).

\bibitem{maru} 
T.~Maruyama et al.,  this proceedings.

\bibitem{Oyamatsu04} 
K.~Oyamatsu et al.,  this proceedings.


\end{thebibliography}
\end{document}